# Thermal conductivity of crystalline AlN and the influence of atomic-scale defects

Runjie (Lily) Xu[1], Miguel Muñoz Rojo[1,2], S. M. Islam[3], Aditya Sood[1,4,5], Bozo Vareskic[6], Ankita Katre[7,8], Natalio Mingo[7], Kenneth E. Goodson[4,9], Huili (Grace) Xing[3,10], Debdeep Jena[3,10] and Eric Pop[1,9,*]

[1]*Electrical Engineering, Stanford University, Stanford, CA 94305, U.S.A.*
[2]*Thermal and Fluid Engineering, University of Twente, Enschede, 7500 AE, Netherlands*
[3]*Electrical & Computer Engineering, Cornell University, Ithaca, NY 14853, U.S.A.*
[4]*Mechanical Engineering, Stanford University, Stanford, CA 94305, U.S.A.*
[5]*Stanford Institute for Materials and Energy Sciences, SLAC National Accelerator Laboratory, Menlo Park, CA 94025, USA*
[6]*Physics and Astronomy, UCLA, Los Angeles, CA 90095, U.S.A.*
[7]*LITEN, CEA-Grenoble, 17 Avenue des Martyrs, 38054 Grenoble, France*
[8]*Centre for Modeling and Simulation (CMS), Savitribai Phule Pune University, Ganeshkhind, Pune-411007, Maharashtra, India*
[9]*Materials Science & Engineering, Stanford University, Stanford, CA 94305, U.S.A.*
[10]*Materials Science & Engineering, Cornell University, Ithaca, NY 14853, U.S.A.*

Aluminum nitride (AlN) plays a key role in modern power electronics and deep-ultraviolet photonics, where an understanding of its thermal properties is essential. Here we measure the thermal conductivity of crystalline AlN by the $3\omega$ method, finding it ranges from $674 \pm 56$ Wm$^{-1}$K$^{-1}$ at 100 K to $186 \pm 7$ Wm$^{-1}$K$^{-1}$ at 400 K, with a value of $237 \pm 6$ Wm$^{-1}$K$^{-1}$ at room temperature. We compare these data with analytical models and first principles calculations, taking into account atomic-scale defects (O, Si, C impurities, and Al vacancies). We find Al vacancies play the greatest role in reducing thermal conductivity because of the largest mass-difference scattering. Modeling also reveals that 10% of heat conduction is contributed by phonons with long mean free paths (MFPs), over ~7 µm at room temperature, and 50% by phonons with MFPs over ~0.3 µm. Consequently, the effective thermal conductivity of AlN is strongly reduced in sub-micron thin films or devices due to phonon-boundary scattering.

*Contact:* epop@stanford.edu

## I. INTRODUCTION

Wide band gap (WBG) semiconductors such as GaN, Ga$_2$O$_3$ and AlN have attracted much interest due to their potential applications in power and radio-frequency (RF) electronics,[1, 2, 3] as well as deep ultraviolet (UV) photonics.[4,5] In these contexts, heat dissipation is important during high-power and high-temperature operation.[6, 7, 8] For example, power devices handle hundreds or even thousands of Volts, and the high power density leads to high operating temperature due to Joule heating, potentially diminishing the device performance and lifetime. Thermal cycling also causes fatigue and eventual failure in such devices.[9, 10]

Among WBG materials, AlN has a large direct band gap (~6.1 eV, almost twice that of SiC and GaN)[11, 12, 13] and one of the largest thermal conductivities. In this respect, as shown in Fig. 1, AlN is among a rare class of materials that have both a large electronic band gap and a large thermal conductivity. AlN is widely used as buffer for GaN growth or as capping layer[14, 15] in power high-electron mobility transistors (HEMTs). However, many questions remain about the role of intrinsic defects and impurities which can occur during AlN growth. The contribution of individual phonon modes to thermal transport in AlN is also not well understood, which is important in establishing the dependence of AlN thermal conductivity on film thickness. (The contribution of electrons to thermal transport is negligible in WBG materials.)

Here, we elucidate these features of AlN thermal transport, by combining 3$\omega$ thermal measurements from 100 to 400 K, with thermal modeling using both analytical and *ab initio* techniques. We uncover that Al vacancies play an important role in limiting the thermal conductivity of present samples, and that phonons with long mean free paths (MFPs > 0.3 μm) contribute over 50% of the thermal conductivity at room temperature. This implies that the effective crystalline AlN thermal conductivity is strongly reduced in submicron films, and could be as low as ~25 Wm$^{-1}$K$^{-1}$ in a 10 nm thin film.

## II. MOTIVATION AND COMPARISON

Figure 1 summarizes the room temperature thermal conductivities of several representative bulk solids with respect to their electronic band gaps. In this plot, a few trends emerge: First, among conducting, zero band gap materials, Cu and graphite (parallel to the basal plane) have the highest thermal conductivity.[16] (Cu is the only material on this plot whose thermal conductivity is dominated by electrons.) Second, among crystalline semiconductors the thermal conductivity weakly scales with the electronic band gap,[17,18,19] as both depend on the strength of the interatomic bonds and (inversely) on the atomic mass. Crystalline boron arsenide (BAs) is somewhat of an exception, with high thermal conductivity despite a relatively moderate electronic band gap, due to its unusual optical-acoustic phononic gap.[20,21] However, polycrystalline and amorphous semiconductors (e.g. poly-Si and a-Si) have much reduced thermal conductivity due to grain boundary and disorder scattering, respectively.[22,23] Third, many electrical insulators, like sapphire, SiO$_2$ or $SiN_x$, have low thermal conductivity.[24,25,26] Thus, only few materials have both large thermal conductivity and large electronic gap, i.e. diamond,[16] hexagonal boron nitride ($h$-BN)[27] (parallel to the basal plane) and AlN, as circled in Fig. 1.

These three materials can provide excellent heat dissipation, especially in power electronics where large amounts of heat are generated. These materials can also be doped, to be used within or as parts of active device regions. The fundamental properties that lead to their high thermal conductivity are small atomic mass, strong inter-atomic bonds, and simple crystal structure. However, the thermal properties of AlN have been studied relatively less[28,29] compared to other WBG materials, and details regarding the role of defects and phonon MFPs, particularly as a function of temperature and sample thickness, are still missing and thus the subject of this work.

## III. MEASUREMENTS AND MODELING

### A. $3\omega$ experimental measurements

The AlN bulk crystals (500 μm thick) were grown using physical vapor transport (PVT).[30] These samples have some imperfections, including Al vacancies and substitutional point defects[31] of oxygen (O), carbon (C) and silicon (Si) atoms, all in the range of $0.4 \times 10^{19}$ to $2 \times 10^{19}$ cm$^{-3}$. Figure 2 shows a schematic of the $3\omega$ setup, which is a method for thermal conductivity measurements using AC-heated electrical lines that also serve as thermometers, well described elsewhere.[25,32,33] Here, four-probe metal lines (5 nm Ti followed by 60 nm Pd) are patterned by optical lithography and lift-off on the AlN sample surface (additional information in Sec. A of Supplementary Material), serving as both heaters and thermometers, as shown in Fig. 2(a). The electrical schematic of the $3\omega$ measurement is displayed in Fig. 2(b).

As shown in Fig. 3(a), an AC current ($I_{1\omega}$) at frequency $\omega$ is passed through the heater, which causes a second harmonic temperature rise ($\Delta T_{2\omega}$) in the sample due to Joule heating. The metal heater line resistance scales linearly with temperature from 100 K to 400 K, as $R = R_0[1 + \alpha(T - T_0)]$, where $\alpha = (5.5 \pm 0.2) \times 10^{-3}$ K$^{-1}$ is the temperature coefficient of resistance (TCR) and $T_0 = 100$ K, as in Fig. 3(b). Due to this linear relationship, the measured line resistance will also have a component ($R_{2\omega}$) that is a second harmonic of the frequency. According to Ohm's Law, the heater output voltage has both $1\omega$ and $3\omega$ components, $V_{1\omega+3\omega} = R_{2\omega}I_{1\omega} = V_{1\omega} + V_{3\omega}$. We use a custom-built circuit board, schematically displayed in Fig. 2(b), to separate $V_{3\omega}$ from $V_{1\omega+3\omega}$.[34] A potentiometer ($R_{variable}$) which has a low TCR of 50 ppm/K is adjusted to match the resistance of the sample heater ($R_{sample}$). When these two resistance values are matched, the voltage drop across the potentiometer is $V_{1\omega}$. Both $V_{1\omega}$ and $V_{1\omega+3\omega}$ are input to a lock-in amplifier, as shown in Fig. 2(b), and $V_{3\omega}$ of the sample is the difference of these two voltage signals.

After collecting the $3\omega$ voltage data, we analytically extract the thermal conductivity of the AlN sample as follows. The $3\omega$ voltage $V_{3\omega}$ vs. frequency $f = \omega/(2\pi)$ is shown in Fig. 3(c). The real part of $V_{3\omega}$ is plotted vs. $\ln(f)$ in Fig. 3(d), displaying a linear variation whose slope $S$ leads to the thermal conductivity $k$ of the sample:

$$k = \frac{R\frac{dR}{dT}I_{1\omega}^3}{4\pi LS},$$ (1)

where $L$ is the length and $R$ is the resistance of the heater, $dR/dT = \alpha R_0$, and $I_{1\omega}$ is the magnitude of the AC current. We used heater dimensions that were 2 mm long (between inner voltage probes) and 20 μm wide, allowing us to treat the heater as a one-dimensional line.[32] Thus, heat flow is perpendicular to the top sample surface, which is in the same direction as the (few) dislocation line defects. The density of dislocation lines provided by the manufacturer[30] is in the range of $10^2$ to $10^4$ cm$^{-2}$, which is expected to have a small impact on the thermal conductivity.[35]

The extracted temperature-dependent thermal conductivities of two single crystal AlN samples are plotted in Fig. 4(a), from 100 K to 400 K. (Sample I shown in red diamonds and sample II shown in blue diamonds.) All measurements were performed in a vacuum probe station (< $10^{-4}$ Torr). As a cross-check, we also used time-domain thermoreflectance (TDTR)[36-38] to measure the thermal conductivity of sample II at room temperature [white diamond in Fig. 4(a)], confirming the accuracy of our measurements. The average thermal conductivity of these AlN samples ranges from 674 ± 56 Wm$^{-1}$K$^{-1}$ at 100 K to 186 ± 7 Wm$^{-1}$K$^{-1}$ at 400 K. At room temperature, the average thermal conductivity is 237 ± 6 Wm$^{-1}$K$^{-1}$ measured by the $3\omega$ method and 247 ± 20 Wm$^{-1}$K$^{-1}$ by TDTR (for sample II), these values being consistent with each other and similar to others reported in the literature.[29,28] We also report the thermal boundary conductance (TBC), $G_b \approx 117$ MWm$^{-2}$K$^{-1}$ at room temperature between AlN and the Al metal pad used in TDTR, with additional details provided in Supplementary Material Section B. The uncertainty due to this TBC during $3\omega$ measurements is negligible due to the large thermal diffusion length at our frequencies (100 to 250 μm) but could play a role in thinner AlN films and devices. (The Kapitza length of AlN corresponding to this TBC is $k/G_b \sim 2.2$ μm at room temperature, meaning that heat flow across AlN films thinner than this value could be partly limited by the thermal resistance of their interfaces, $1/G_b$.)

**B. Analytical model**

To analyze the contributions of different phonons and understand the underlying phonon scattering mechanisms in AlN, we turn to computational modeling, using two approaches: (1) we fit the measured data to an analytical model based on the Boltzmann Transport Equation (BTE), and (2) we perform full ab initio calculations. The analytical model [black solid line in Fig. 4(a)] is calculated based on the simplified

BTE, using the Debye approximation for the phonon dispersion of the acoustic modes (additional details are in the Supplementary Material Section D):[35, 39]

$$k = \frac{1}{3} C v \lambda = \frac{1}{3} \sum_s \int_0^{\omega_{max}} \hbar \omega g(\omega) \frac{df(\omega,T)}{dT} v^2 \tau(\omega) d\omega \qquad , \qquad (2)$$

where $\lambda$ is the phonon MFP, $v$ is the phonon group velocity, $C$ is the heat capacity, $\omega$ is the phonon frequency, $\omega_{max}$ is the Debye cutoff frequency, $g(\omega)$ is the phonon density of states, $f(\omega,T)$ is the Bose-Einstein distribution, $\tau(\omega)$ is the phonon scattering time, and $s$ includes two transverse acoustic (TA) phonon modes and one longitudinal acoustic (LA) mode of AlN. The scattering rate is

$$\frac{1}{\tau} = \frac{1}{\tau_N} + \frac{1}{\tau_U} + \frac{1}{\tau_D} + \frac{1}{\tau_B} \quad , \qquad (3)$$

where the subscripts correspond to normal-process (N), Umklapp (U), defect (D), and boundary (B) scattering, respectively. Point defect scattering arises from impurity atoms of C, Si, and O, and from Al vacancies. As it turns out, the latter plays an important role in the reduction of thermal conductivity in this study, and the point defect scattering rate can be written as[40]

$$\frac{1}{\tau_D} = \frac{V}{4\pi v^3} \omega^4 \sum_i f_i (\frac{m-m_i}{m})^2, \qquad (4)$$

where $V$ is the unit volume for wurtzite AlN given by $V = \sqrt{3}a^2c/8$, $a = 3.11$ Å and $c = 4.98$ Å are lattice constants,[41] $f_i$ is the fractional concentration of the $i$-th impurity atom, $m$ and $m_i$ are the masses of original and $i$-th impurity atoms, respectively. In point defect scattering, Al vacancies play a dominant role because the mass difference is the atomic mass of the Al atom, which is much larger than the mass difference between Si and Al atoms or the difference among O, C, and N atoms. In AlN, C atoms often substitute for N atoms, while Si substitutes for Al.[31] In our analytical model, the Al vacancy density is used as a fitting parameter, with a fitted value of ~2×10[19] cm[-3], which is within the range quoted by the sample manufacturer.[30] An important "shortcut" used here for treating vacancy scattering relies on a previous study by Katcho *et al.*[42] which showed good agreement with first principles calculations if the vacancy mass difference is taken as six times the mass of the missing atom. This is justified because vacancies lead to larger local distortion in the crystal compared to substitutional defects, due to bond breaking and atomic rearrangements, and these distortions contribute to enhanced phonon scattering.

## C. First principles calculations

We also employ a second modeling approach, using first principles calculations, based on the BTE coupled with density functional theory (DFT). This method has previously shown good agreement with experiments for a range of other materials.[43,44,45] The phonon frequencies and anharmonic phonon scattering rates for AlN are computed using harmonic (2nd order) and anharmonic (3rd order) interatomic force constants (IFCs) for a 5×5×5 supercell of AlN wurtzite structure (space group P6₃mc). We follow the finite displacement method as implemented in phonopy[46] and thirdorder.py,[47] extracting the 2nd and 3rd order IFCs respectively from interatomic forces. These interatomic forces and the optimized structural parameters for wurtzite AlN are calculated using the DFT package VASP,[48] and additional details are provided in the Supplementary Material Section E. Similar to the analytic approach described earlier, the phonon scattering rate with Al vacancies is computed using Eq. 4, where the mass difference is six times the original atomic mass.[42] All contributions to phonon scattering rates and finally the thermal conductivity are calculated using the almaBTE package,[49] where the BTE is solved using an iterative scheme, and the obtained thermal conductivity is shown with a purple dashed line in Fig. 4(a), displaying good agreement with the experiments.

We note that the analytic and first-principles calculations fit the thermal conductivity data with different Al vacancy concentrations, i.e. $2 \times 10^{19}$ cm⁻³ and $4 \times 10^{18}$ cm⁻³, respectively, although both are in the range quoted by the sample manufacturer.[30] This difference is due to the different anharmonic scattering rates implemented in the two approaches. In the analytical model, anharmonic scattering rates for both normal and Umklapp processes follow the simple $\omega^2$ behavior.[28] The anharmonic scattering rates in the ab initio calculations show deviation from this behavior at both low and high frequencies.[44, 50] However, we note that the five-fold difference in vacancy concentration causes only about ~25% change of expected bulk thermal conductivity [Fig. 4(b)], illustrating the relative (in)sensitivity of this parameter in this range.

## IV. THICKNESS DEPENDENCE OF THERMAL CONDUCTIVITY

Figure 4(b) examines the AlN thermal conductivity dependence on vacancy concentration and film thickness. The thickness dependence with different vacancy concentrations has not been previously analyzed before, although (as we will see) AlN is subject to strong phonon-boundary scattering effects due to the large phonon MFP in this material. In other words, the thermal conductivity of sub-micron thin AlN films is strongly reduced, and thin buffer films of this material are expected to have much lower effective thermal conductivity than the bulk material. This is an *intrinsic* effect, in addition to the earlier observation of *extrinsic* thermal impedance contribution from interfaces (like Al/AlN) of sub-micron thin films.

Figure 4(b) displays the calculated thickness-dependent thermal conductivity with different defect densities using solid lines, all at room temperature. For comparison, experimental data on various single crystal films are shown in diamond symbols, including this work and Refs. [29,51,52]. Square symbols correspond to one bulk polycrystalline AlN measured with TDTR[53] and other polycrystalline films measured by various groups.[52,54,55,56,57,58,59] Round symbols correspond to amorphous thin films by Zhao *et al.*[55] and Gaskins *et al.*[60] Due to significant disorder scattering, amorphous films have much lower thermal conductivity than (poly-)crystalline films, as expected. Thus, when using AlN thin films as buffer or capping layers[14,15] in power devices, highly crystalline, low-defect films provide better heat dissipation.

However, Fig. 4(b) also reveals that the thermal conductivity of all films ~10 μm or thinner is expected to be decreased by ~10% or more from the bulk value. The effective thermal conductivities of 10 nm and 100 nm thin AlN films are predicted to be just ~25 Wm⁻¹K⁻¹ and ~110 Wm⁻¹K⁻¹ at room temperature (less than 1/12 and 1/3 of the best bulk material values), respectively, even in defect-free films, due to strong phonon-boundary scattering.

## V. ACCUMULATED THERMAL CONDUCTIVITY

To understand the physical origin of the strong phonon-boundary scattering in AlN thin films, we turn to Fig. 5. First, in Fig. 5(a) we plot the calculated thermal conductivity as a function of the cumulative contributions of phonons across the range of MFPs expected in such crystals. The accumulated thermal conductivity is the thermal conductivity contribution from all phonons with MFP below a given value:[61]

$$k_{\text{accum}}(\lambda_0) = \frac{1}{3} \sum_s \int_0^{\lambda_0} C(\lambda)\, v(\lambda) \lambda d\lambda, \tag{5}$$

where $C$ is the heat capacity as a function of MFP, since $C(\omega) = \hbar \omega g(\omega)\, df(\omega, T)/dT$ and $\lambda = v\tau(\omega)$. The integral is taken from 0 to $\lambda_0$, and thus $k_{\text{accum}}$ is the thermal conductivity of phonons with MFP $\leq \lambda_0$, here at room temperature. The contributions of both LA and TA modes are shown in Fig. 5(a), the LA mode contribution being larger due to its larger phonon group velocity. The total thermal conductivity is the sum of contributions from one LA and two TA modes.

To gain additional insight, we normalize the accumulated thermal conductivity by the bulk value ($k_{\text{accum}}/k_{\text{bulk}}$) in Fig. 5(b), for the "perfect crystal" with zero defects. Our calculations estimate that 50% of the AlN bulk thermal conductivity is contributed by phonons with MFPs > 0.3 μm, and 10% is contributed by phonons with very long MFPs > 7 μm, at room temperature. These values are comparable to the median MFP ~ 2.5 μm of Freedman *et al.*,[61] obtained by broadband frequency domain thermoreflectance (BB-FDTR) which considered only Umklapp phonon scattering (vs. the four scattering mechanisms included here). Taken together, these findings explain why "size effects" on the thermal conductivity of AlN are expected to be strong in sub-micron films at room temperature, and noticeable even in sub-10 μm thin films. In other words, the effective thermal conductivity of AlN is strongly reduced in films with thickness comparable to or smaller than such long phonon MFPs, as illustrated earlier in Fig. 4(b).

We define the phonon MFP corresponding to 50% or 90% of the cumulative heat conduction as MFP(50% or 90%), plotting it at higher temperatures in Fig. 5(c). As the temperature increases, phonon occupation and phonon-phonon scattering increase, thus MFP(50% or 90%) decreases. This implies that "size effects" on the thermal conductivity of AlN become somewhat less important at elevated temperature, i.e. the reduction of thermal conductivity in thin films of this material will be less pronounced vs. the bulk value at that temperature. The thermal conductivity of thin films at high temperatures will also experience a competition between phonon-phonon and phonon-boundary scattering. This is illustrated in Fig. 5(d), which shows the expected temperature dependence of thermal conductivity from bulk to 1 μm, 0.1 μm, and 10 nm thin films. The increasing role of phonon-boundary scattering lowers the thermal conductivity, but also renders it less temperature-sensitive in the thinnest films, and less dependent on film thickness at the highest temperatures. The exact details of boundary scattering processes will depend, in part, on the particular surface roughness of such AlN films. These details were previously studied for Si, Ge and GaAs thin films and nanowires,[62,63] and should be the subject of future work for AlN.

## VI. CONCLUSIONS

In summary, we have performed $3\omega$ measurements of thermal conductivity in single crystal AlN samples, from 100 K to 400 K. We compared these results with analytic and ab initio simulations, to estimate the impurity defect densities. Aluminum vacancies play the most important role among all atomic scale defects, due to the large atomic mass mismatch, which can be analytically captured by modeling phonon-vacancy scattering using six times the mass of the missing atom. The accumulated thermal conductivity shows that phonons with MPFs larger than 0.3 μm (or 7 μm) contribute to 50% (or 10%) of heat conduction at room temperature. This implies that AlN thin films and devices with sub-micron features will exhibit strongly reduced effective thermal conductivity compared to the bulk value, even in the absence of point defects. These results are essential for the understanding of thermal transport in AlN thin films and devices over a broad temperature range, for applications in power electronics and deep-UV lasers.

## SUPPLEMENTARY MATERIAL

See supplementary material for the additional details of the fabrication process, TDTR measurement, analytical model and first principles calculations.

## ACKNOWLEDGMENTS

This work was supported in part by the National Science Foundation (NSF) DMREF program grants 1534279 and 1534303, by the NSF Engineering Research Center for Power Optimization of Electro-Thermal Systems (POETS) with cooperative agreement EEC-1449548, and by the Stanford SystemX Alliance. This work was also supported by ASCENT, one of six centers in JUMP, a SRC program sponsored by DARPA. Work was performed in part at the Stanford Nanofabrication Facility and the Stanford Nano Shared Facilities which receive funding from the NSF as part of the National Nanotechnology Coordinated

Infrastructure award ECCS-1542152. A.K. and N.M. acknowledge support from the Air Force Office of Scientific Research grant FA9550615-1-0187 DEF. A.K. also acknowledges DST-INSPIRE Grant, India (Grant No. IFA17-MS122). R.L.X. and M.M.R. gratefully acknowledge technical discussions with C. Dames, V. Mishra and W. Hodges.

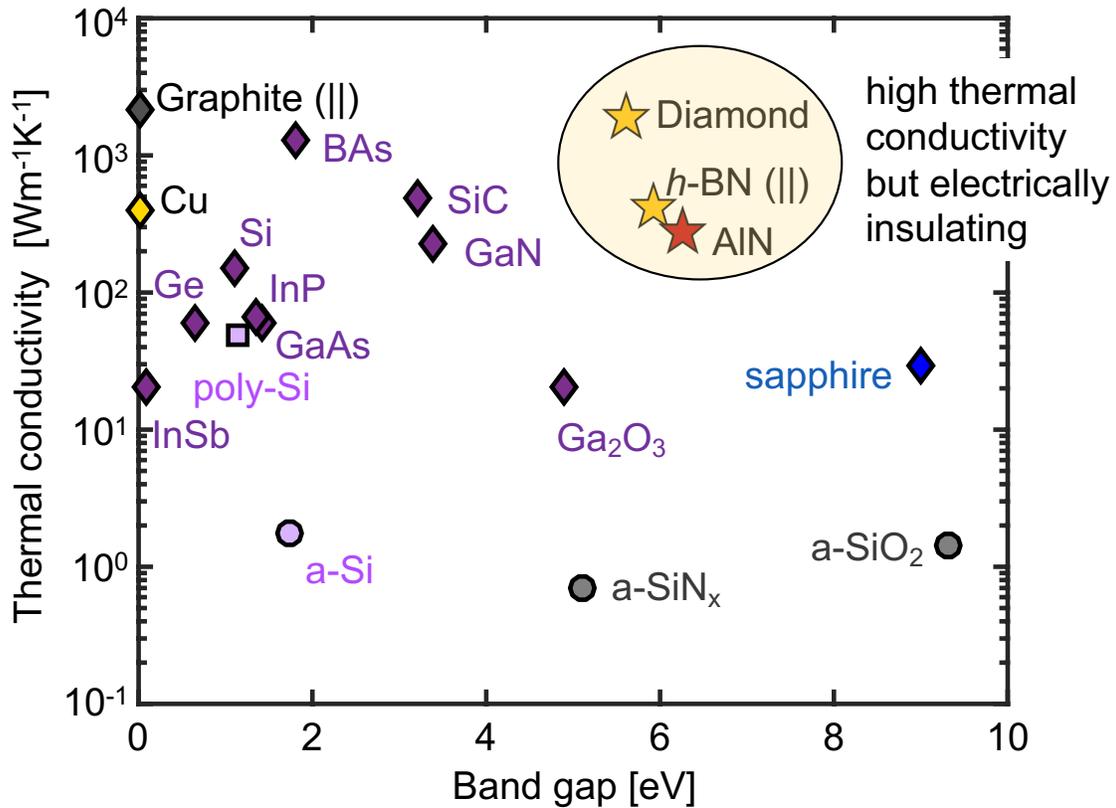

**Figure 1.** Room temperature thermal conductivities of different materials vs. their electronic band gaps. These include electrical conductors (e.g., graphite and Cu),[16] semiconductors (e.g., Si,[16,22,23] Ge,[16] InSb,[17] InP,[64] GaAs,[17] BAs,[20,21] SiC,[19] GaN,[65,66] and Ga₂O₃[67]), and some electrical insulators (e.g., diamond,[16] *h*-BN,[27] AlN,[28,29] sapphire,[24] amorphous SiO₂[25] and amorphous SiNₓ[26]). The plot reveals that AlN lies in the same range as diamond and *h*-BN (star symbols), with both wide band gaps and high thermal conductivities. Isotopically purified samples may have higher thermal conductivity (values displayed are for natural isotopes). Diamonds are for crystalline, squares for polycrystalline, and circles for amorphous materials.

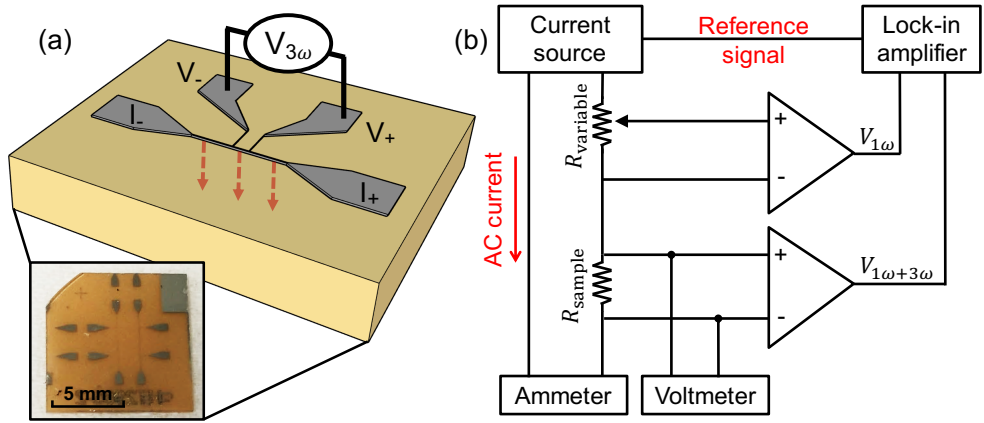

**Figure 2.** (a) Schematic of four-probe $3\omega$ metal heater line on AlN single crystal sample. Heater consists of 5 nm Ti and 60 nm Pd, 20 μm wide and 2 mm long between the inner voltage probes. Arrows indicate heat flow direction. Inset shows an optical image of one of the AlN samples with patterned $3\omega$ heaters. (b) Electronic circuit and instrument setup of the $3\omega$ measurement.

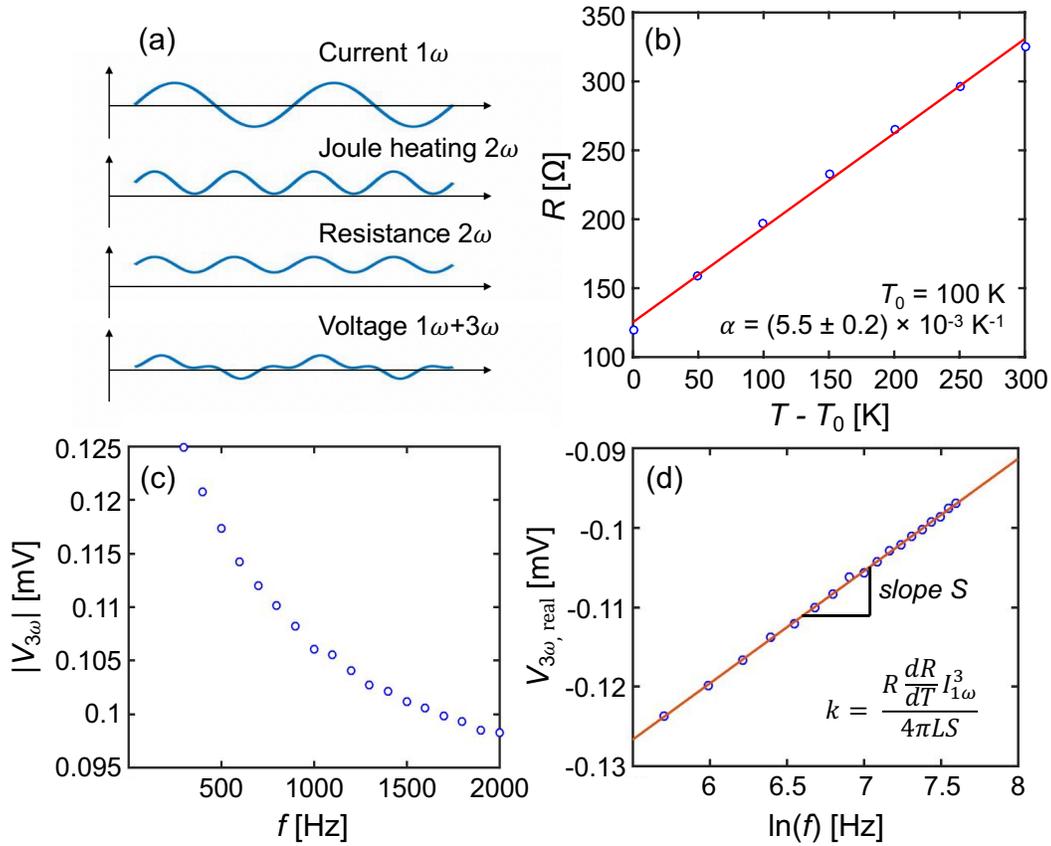

**Figure 3.** Analysis of 3$\omega$ measurement. (a) An AC current of frequency 1$\omega$ is passed through the heater line. Joule heating causes a second harmonic temperature rise, at 2$\omega$, in the AlN sample underneath the heater. The metal heater resistance varies linearly with temperature as $R = R_0[1 + \alpha(T - T_0)]$, where $\alpha$ is the TCR and $T_0$ is the background temperature. Due to this linear relationship, the measured heater resistance will also have a 2$\omega$ component dependent on the sample temperature. Multiplied by the AC current input, the output voltage will have a component at 3$\omega$. (b) TCR measurement fitting of sample I. (Sample II data are shown in the Supplementary Material Fig. S1.) Symbols are experimental data, solid line is the fit. (c) Measured $|V_{3\omega}|$ vs. frequency $f$. The real part of $V_{3\omega}$ is linear with ln($f$), as shown in (d). Blue circles are measured data, and the thermal conductivity $k$ is calculated using the slope of the linear fit (solid line).

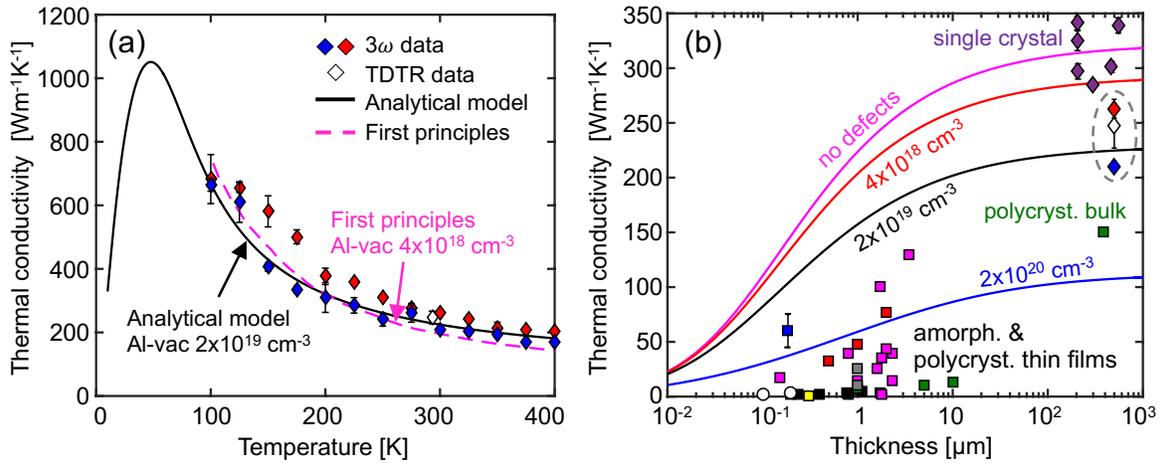

**Figure 4.** (a) Thermal conductivity of AlN vs. temperature. Square symbols are experimental data measured by our 3ω method. Diamond symbol is measured using TDTR. Dashed line is the model calculated by first principles simulation. Solid line is the thermal conductivity calculated by the analytical model. (b) Thermal conductivity of AlN vs. sample thickness, at room temperature. Solid lines are the theoretical calculation using different AlN defect densities. Diamond symbols are single crystal samples measured in this work [circled, colors matching panel (a)], those by Slack *et al.*[28], and Rounds *et al.*[29] Square symbols are a poly-crystalline bulk sample[53] (in green) and various polycrystalline films (grey: Kuo *et al.*[52], purple: Duquenne *et al.*[54], black: Zhao *et al.*[55], red: Choi *et al.*[56], blue: Yalon *et al.*[57], yellow: Jacquot[58] *et al.*, green: Bian *et al.*[59]). White round symbols correspond to amorphous thin films by Zhao *et al.*[55] and Gaskins *et al.*[60]

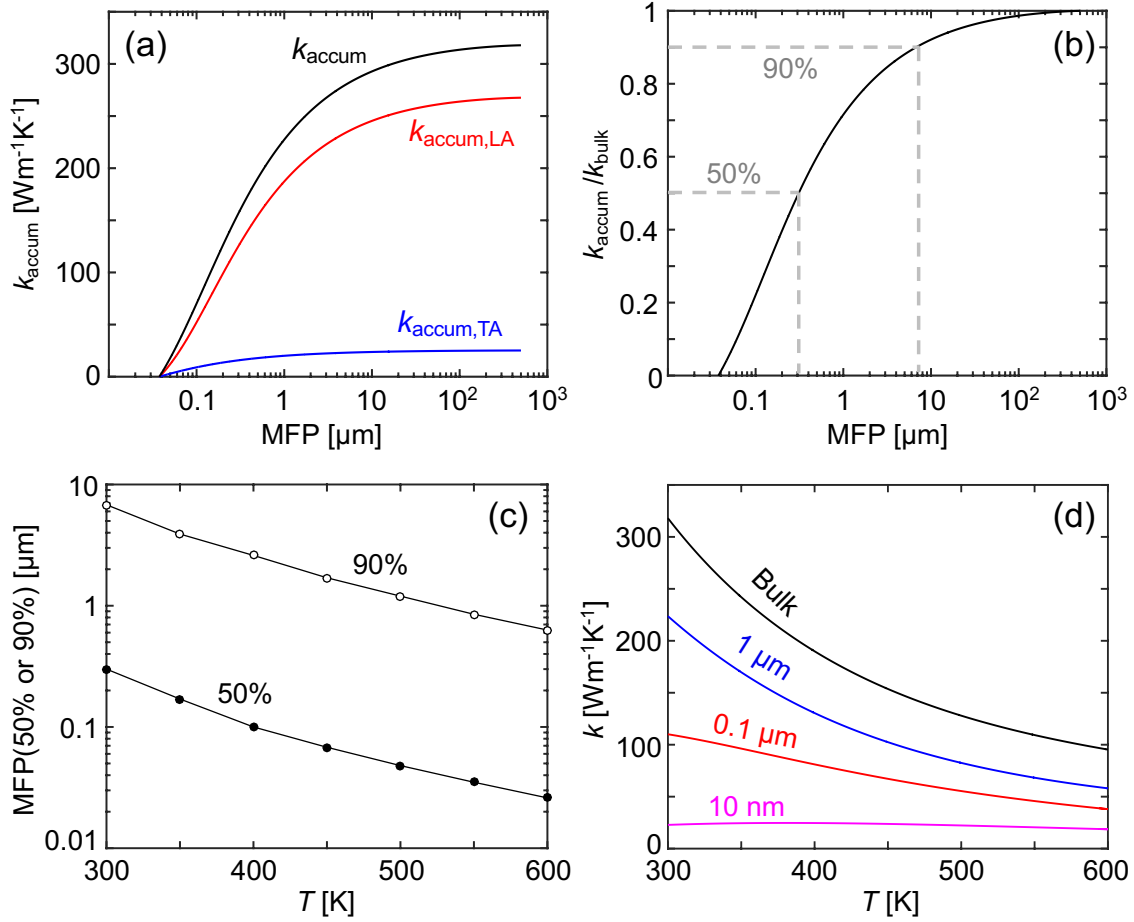

**Figure 5**. (a) Calculated accumulated thermal conductivity vs. phonon MFP for AlN bulk at room temperature, comparing the total and its longitudinal acoustic and transverse acoustic phonon contributions, $k_{accum}$ = $k_{accum,LA}$ + $2k_{accum,TA}$. (b) Normalized accumulated thermal conductivity $k_{accum}$ / $k_{bulk}$ at room temperature, where $k_{bulk}$ is the maximum value of $k_{accum}$. Phonons with MFP larger than 0.3 μm (or 7 μm) are estimated to contribute 50% (or 10%) of the heat conduction, as shown by dashed lines. (c) Calculated temperature dependence of MFP(50% or 90%) for AlN. (d) Expected temperature dependence of thermal conductivity for different film thicknesses, as labeled. Thinner films have weaker temperature dependence, due to the predominance of boundary scattering. All calculations (a-d) in this figure assume defect-free samples.

# Supplementary Information

# Thermal conductivity of crystalline AlN and the influence of atomic-scale defects

Runjie (Lily) Xu[1], Miguel Muñoz Rojo[1,2], S. M. Islam[3], Aditya Sood[1,4,5], Bozo Vareskic[6], Ankita Katre[7,8], Natalio Mingo[7], Kenneth E. Goodson[4,9], Huili (Grace) Xing[3,10], Debdeep Jena[3,10] and Eric Pop[1,9,*]

[1]*Electrical Engineering, Stanford University, Stanford, CA 94305, U.S.A.*

[2]*Thermal and Fluid Engineering, University of Twente, Enschede, 7500 AE, Netherlands*

[3]*Electrical & Computer Engineering, Cornell University, Ithaca, NY 14853, U.S.A.*

[4]*Mechanical Engineering, Stanford University, Stanford, CA 94305, U.S.A.*

[5]*Stanford Institute for Materials and Energy Sciences, SLAC National Accelerator Laboratory, Menlo Park, CA 94025, USA*

[6]*Physics and Astronomy, UCLA, Los Angeles, CA 90095, U.S.A.*

[7]*LITEN, CEA-Grenoble, 17 Avenue des Martyrs, 38054 Grenoble, France*

[8]*Centre for Modeling and Simulation (CMS), Savitribai Phule Pune University, Ganeshkhind, Pune-411007, Maharashtra, India*

[9]*Materials Science & Engineering, Stanford University, Stanford, CA 94305, U.S.A.*

[10]*Materials Science & Engineering, Cornell University, Ithaca, NY 14853, U.S.A.*

[*]**Contact:** epop@stanford.edu

## A. Heater Line Fabrication Details and TCR

The $3\omega$ heater lines used in this work are 20 μm wide and 2 mm long (between the inner voltage probes, see Fig. 2 in the main text). The lines were fabricated using photo-lithography. First, the AlN sample was spin-coated with a lift-off layer, LOL 2000, at 3000 rpm for 60 seconds, and pre-baked at 170°C for 7 minutes. Then, a photoresist, SPR 3612, was spin-coated on top at 5500 rpm for 40 s and pre-baked at 90°C for 60 seconds. A photomask was used to pattern the heater lines after 3 s exposure with an ultraviolet (UV) mask aligner lithography tool (Karl Suss). Afterwards, we carried out a post-exposure bake at 115°C for 60 s. We developed our sample by immersing it in MF-26A developer for 60 s. This was followed up by a soaking process in distilled water for 1 min and a gentle dry with compressed air. Then, we evaporated 5 nm Ti and 60 nm Pd using an e-beam evaporator (KJ-Lesker). Finally, the lift-off process was done by soaking the sample for 20 min in remover PG at 60°C.

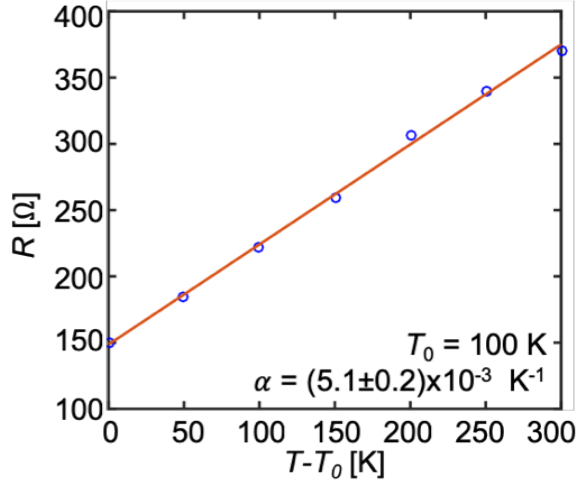

**Fig. S1.** Measurement (symbols) and fitting (line) of temperature coefficient of resistance (TCR) for Sample II. Here $R = R_0[1 + \alpha(T - T_0)]$, where $T_0 = 100$ K and $R_0$ is measured at $T_0$. The TCR measurement of Sample I is shown in Fig. 3(b) of the main text.

## B. Time-Domain Thermoreflectance (TDTR)

TDTR measurements were performed at room temperature to compare with $3\omega$ results. Details of this technique and our setup are provided elsewhere.[1] An 80 nm thick Al layer was deposited on the AlN crystal by electron-beam (e-beam) evaporation. Measurements were made at a modulation frequency of 10 MHz, with pump and probe spot sizes ($1/e^2$ diameters) of 10.2 and 6.2 μm, respectively. The incident powers of pump and probe lasers beams were ~15 and 3 mW, respectively.

TDTR ratio data were fit to the solution of a three-dimensional (3D) heat diffusion model.[2] Al thickness was measured using atomic force microscopy (= 80 ± 2 nm). The heat capacity of single-crystal AlN at room temperature was taken from literature (= 2.4 MJm$^{-3}$K$^{-1}$).[3] Two unknown properties were extracted through a simultaneous fit: the AlN thermal conductivity (= 247 ± 20 Wm$^{-1}$K$^{-1}$) and the thermal boundary conductance (TBC) between the Al and AlN (= 117 MWm$^{-2}$K$^{-1}$). This TBC corresponds to a thermal interface resistance TBR = 1/TBC ≈ 8.55 m$^2$K/GW, all at room temperature. The data and best fit result are shown in Fig. S2.

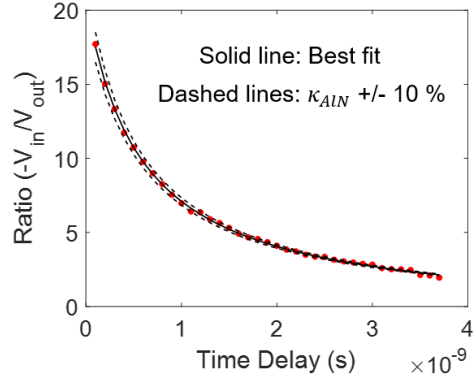

**Fig. S2.** TDTR data (red circles) and best fit result (solid line). Dashed lines represent perturbations of ±10% in the AlN thermal conductivity with respect to the best fit value.

### C. Calibration Samples and Uncertainty Analysis

We have used a fused quartz wafer as calibration sample of our $3\omega$ setup. Temperature rise ($\Delta T$) *vs.* driving frequency on fused quartz is shown in Fig. S3. The thermal conductivity of fused quartz was extracted as 1.5 Wm$^{-1}$K$^{-1}$ at room temperature, which is within 7% of a previous study by Abdulagatov *et al.*[4] We used the same heater design for the calibration sample and our AlN samples, as detailed in Section A.

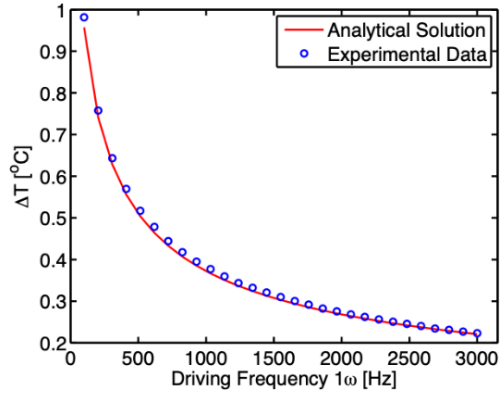

**Fig. S3.** Calibration of $3\omega$ thermometry setup. Temperature rise *vs.* driving frequency on fused quartz wafer. Blue circles are measured data and red curve is our analytical model.

Error bars in the reported thermal conductivity values measured by $3\omega$ are calculated by error propagation. The uncertainty value of thermal conductivity $\delta k$ is calculated using the uncertainties in heater length $\delta L$ (~5 μm i.e. 0.25% out of $L = 2$ mm), ac current value $\delta I$ (~10 μA), heater resistance $\delta R$ (from 0.4 Ω at 100 K to 0.2 Ω at 400 K), slope of $V_{3\omega}$ *vs.* ln($f$), $\delta S$ (from $1.9 \times 10^{-7}$ at 100 K to $6.7 \times 10^{-7}$ at 400 K) and fitting error from TCR measurements d$R$/d$T$ (± 0.03 Ω/K, from TCR in Section A).

$$\frac{\delta k}{k} = \sqrt{\left(\frac{\delta R}{R}\right)^2 + \left(\frac{\delta \frac{dR}{dT}}{\frac{dR}{dT}}\right)^2 + 3\left(\frac{\delta I}{I}\right)^2 + \left(\frac{\delta L}{L}\right)^2 + \left(\frac{\delta S}{S}\right)^2} \qquad \text{(S.1)}$$

Error bars in the reported TDTR values are calculated by propagating uncertainties in the Al transducer thickness (± 2 nm) and root mean square (rms) laser spot size (± 5 %). These translate respectively to uncertainties of approximately ± 5 % and ± 6 % in the thermal conductivity. The total uncertainty is obtained by adding up these individual contributions in quadrature (± 8 %).

## D. Analytical Model Calculation Details

The thermal conductivity $k$ of AlN is calculated using the Boltzmann Transport Equation (BTE), and the Debye approximation for the phonon dispersion, as follows:

$$k = \frac{1}{3} C v \lambda = \frac{1}{3} \sum_s \int_0^{\omega_{max}} \hbar \omega g(\omega) \frac{df(\omega, T)}{dT} v^2 \tau(\omega) d\omega \qquad (S.2)$$

where $\lambda$ is the phonon mean free path (MFP), $v$ is the phonon group velocity, $T$ is temperature, $C$ is the heat capacity, $\omega$ is the phonon frequency, $\omega_{max}$ is the Debye cutoff frequency, and the $s$ subscript corresponds to the phonon modes (two transverse, one longitudinal). Transverse acoustic phonons have group velocity $v_{TA} = 5250$ m/s and cutoff $\omega_{max,TA} = 33$ rad/ps. Longitudinal phonons have group velocity $v_{LA} = 10504$ m/s and cutoff frequency $\omega_{max,LA} = 66$ rad/ps, from Fig. S4, along the Γ to A (c-axis) direction. In Equation S.2, $g(\omega)$ is the phonon density of states, $f(\omega, T)$ is the Bose-Einstein distribution, and $\tau(\omega)$ is the phonon scattering time. The scattering rate is

$$\frac{1}{\tau} = \frac{1}{\tau_N} + \frac{1}{\tau_U} + \frac{1}{\tau_D} + \frac{1}{\tau_B} \qquad (S.3)$$

where the subscripts correspond to normal-process (N), Umklapp (U), defect (D), and boundary (B) scattering, respectively. Normal-process scattering time is[5]

$$\frac{1}{\tau_N} = \frac{k_B^3 \gamma^2 V}{M \hbar^2 v^5} \omega^2 T^3 \qquad (S.4)$$

where $k_B$ is the Boltzmann constant, $\hbar$ is the reduced Planck constant, $M$ is the average mass of an atom in the crystal, $\gamma$ is the Grüneisen parameter,[6] $V$ is the unit volume for wurtzite AlN given by $V = \sqrt{3}a^2c/8$, $a = 3.11$ Å and $c = 4.98$ Å are lattice constants.[7] The Umklapp scattering time is

$$\frac{1}{\tau_U} = \frac{\hbar \gamma^2}{M \Theta v^2} e^{\left(-\frac{\Theta}{3T}\right)} \omega^2 T \qquad (S.5)$$

where $\Theta$ is the Debye temperature (950 K).[8] Point defect scattering time is $\frac{1}{\tau_D} = \frac{V}{4\pi v^3} \omega^4 \sum_i f_i \left(\frac{m - m_i}{m}\right)^2$, explained in the main text. Boundary scattering $\frac{1}{\tau_B} = \frac{d}{v}$,[9] where $d = 500$ μm is the thickness of the sample, which is treated as an independent variable in Fig. 4(b) of the main text.

## E. First Principles Computational Details

The calculations for AlN are performed using the projector-augmented-wave method[10] implemented in VASP[11] with the local density approximation (LDA) for exchange-correlation function.[12] First, the AlN wurtzite cell is relaxed to get the optimized structural parameters. The resulting relaxed structure is shown in Fig. S4, where the cell parameters are $a = 3.09$ Å and $c = 4.93$ Å and the wurtzite parameter is $u = 0.38$ Å. These are slightly different from the values we used in the analytical model.

The 2$^{nd}$ and 3$^{rd}$ order interatomic force constants are then calculated for a 5×5×5 supercell (300 atoms) of the hexagonal primitive cell. Non-analytical term correction is also included to reproduce LO-TO (Longitudinal Optic and Transverse Optic) phonon splitting in AlN, for which we have computed the Born effective charges and dielectric tensor using VASP.[13] The calculated phonon dispersion curve and the compari-son with experimental data[14] for AlN are shown in Fig. S5. Scattering rates corresponding to phonon-phonon as well as phonon-defect interactions, and the thermal conductivity is finally calculated using a uniformly spaced converged **q**-point mesh of 26×26×14 with the almaBTE package.[15]

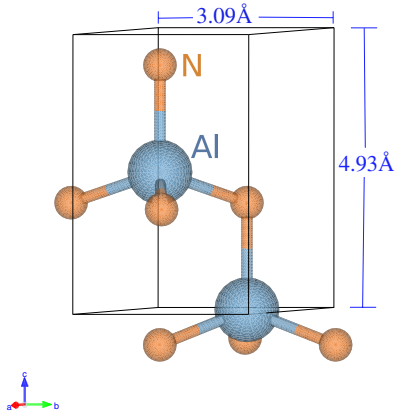

**Fig. S4.** Optimized wurtzite structure of AlN (P6$_3$mc space group). The primitive cell consists of two Al atoms and two N atoms. (More N atoms are seen here due to periodic boundary conditions.)

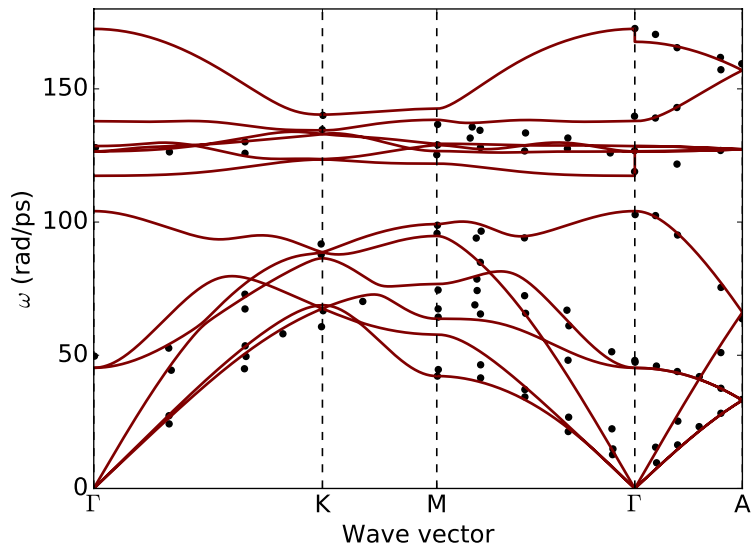

**Fig. S5.** Calculated phonon dispersion for AlN (lines) shown along the high-symmetry Brillouin zone directions, compared with the experimental results of Schwoerer-Böhning *et al.*[14] (symbols).